\documentclass[aoas,MSNbibl,nameyear,dvips]{arximspdf}
\usepackage{graphicx}

\doi{10.1214/12-AOAS600} 
\volume{7}
\issue{1}
\pubyear{2013}
\firstpage{154}
\lastpage{176}

\makeatletter
\newcommand{\rrvert}{\vert}
\newcommand{\llvert}{\vert}
\newcommand{\btheta}{\bolds{\theta}}
\newcommand{\bmu}{\bolds{\mu}}

\newcommand{\bzero}{\mathbf{0}}

\newcommand{\bs}{\mathbf{s}}
\newcommand{\bv}{\mathbf{v}}
\newcommand{\bx}{\mathbf{x}}
\newcommand{\bY}{\mathbf{Y}}
\newcommand{\bZ}{\mathbf{Z}}

\newcommand{\given}{ | }
\newcommand{\bbeta}{ \bolds{\beta}}
\newcommand{\balpha}{ \bolds{\alpha}}
\newcommand{\bphi}{\bolds{\phi}}
\newcommand{\sig}{\sigma}
\newcommand{\ind}{\stackrel{{\mathrm{ind}}}{\sim}}
\newcommand{\iid}{\stackrel{{\mathrm{i.i.d.}}}{\sim}}
\makeatother

\begin{document}
\begin{frontmatter}

\title{Modeling temporal gradients in regionally
aggregated California asthma hospitalization data}
\runtitle{Temporal gradients in regionally aggregated data}

\begin{aug}
\author[A]{\fnms{Harrison}~\snm{Quick}\corref{}\ead[label=e1]{quic0038@umn.com}},
\author[A]{\fnms{Sudipto}~\snm{Banerjee}\ead[label=e2]{sudiptob@biostat.umn.edu}}
\and
\author[A]{\fnms{Bradley P.}~\snm{Carlin}\ead[label=e3]{brad@biostat.umn.edu}}
\runauthor{H. Quick, S. Banerjee and B.~P. Carlin}
\affiliation{University of Minnesota}
\address[A]{Division of Biostatistics\\
University of Minnesota\\
Minneapolis, Minnesota 55425\\
USA\\
\printead{e1}} 
\end{aug}

\received{\smonth{11} \syear{2011}}
\revised{\smonth{9} \syear{2012}}

%
\begin{abstract}
Advances in Geographical Information Systems (GIS) have led to the
enormous recent
burgeoning of spatial-temporal databases and associated statistical modeling.
Here we depart from the rather rich literature in space--time
modeling by considering the setting where space is discrete (e.g.,
aggregated data over regions), but time is continuous.
Our major objective in this application is to carry out inference on
gradients of a temporal process in our data set of monthly
county level asthma hospitalization rates in the state of California,
while at the same time accounting for
spatial similarities of the temporal process across neighboring counties.
Use of continuous time models here allows inference at a finer
resolution than at which the data are sampled.
Rather than use
parametric forms to model time, we opt for a more flexible stochastic
process embedded within a dynamic Markov random field
framework. Through the {matrix-valued covariance function}
we can ensure that the
temporal process realizations are mean square differentiable, and may thus
carry out inference on temporal gradients in a posterior
predictive fashion. We use this approach to evaluate temporal gradients
where we are concerned with temporal changes in the residual and fitted rate
curves after accounting for seasonality, spatiotemporal ozone levels
and several
spatially-resolved important sociodemographic covariates.
\end{abstract}

%

\end{frontmatter}

\section{Introduction}\label{Intro}

Technological advances in spatially-enabled sensor networks and
geospatial information storage, analysis and distribution
systems have led to a burgeoning of spatial-temporal databases.
Accounting for associations across space and time {constitutes} a
routine component in analyzing geographically and temporally referenced
data sets. The inference garnered through these analyses often supports
decisions with important scientific implications, and it is therefore
critical to accurately assess inferential uncertainty. The obstacle for
researchers is increasingly not access to the right data, but rather
implementing appropriate statistical methods and software.

There is a considerable literature in spatio-temporal modeling; see,
for example, the recent book by \citet{CreWik11} and the
references therein. Space--time modeling can broadly be classified as
considering one of the following four settings: (a)~space is viewed as
continuous, but time is taken to be discrete, (b)~space and time are
both continuous, (c)~space and time are both discrete, and (d)~space is
viewed as discrete, but time is taken to be continuous. Almost
exclusively, the existing literature considers the first three
settings. Perhaps the most pervasive case is the first. Here, the data
are regarded as a time series of spatial process realizations. Early
approaches include the STARMA [Pfeifer and Deutsch (\citeyear{PfeDeu80N1,PfeDeu80N2})] and
STARMAX [\citet{Sto86}] models, which add spatial covariance structure
to standard time series models. \citet{HanWal94} employ
stationary Gaussian process models with an $\operatorname{AR}(1)$ model for the time
series at each location to study global warming.
Building upon previous work in the setting of dynamic models by \citet{WesHar97},
several authors, including Stroud, M{\"u}ller and Sans{\'o} (\citeyear{StrMulSan01})
and Gelfand, Banerjee and Gamerman (\citeyear{GelBanGam05}), proposed dynamic frameworks to model
residual spatial and temporal dependence.

When space and time are both viewed as continuous, the preferred
approach is to construct stochastic processes using space--time
covariance functions. \citet{Gne02} built upon earlier work by
\citet{CreHua99} to propose general classes of nonseparable,
stationary covariance functions that allow for space--time interaction
terms for spatiotemporal random processes. \citet{Ste05} considered a
variety of properties of space--time covariance functions and how these
were related to process spatial-temporal interactions.

Finally, in settings where both space and time are discrete there has
been much spatiotemporal modeling based on a Markov random field (MRF)
structure in the form of conditionally autoregressive (CAR)
specifications. See, for example, \citet{Waletal97}, who developed
such models in the service of disease mapping, and Gelfand et al. (\citeyear{Ge98}),
whose interest was in single family home sales. \citet{Pacetal00} work with simultaneous autoregressive (SAR) models extending
them to allow temporal neighbors as well as spatial neighbors. %
Later examples include the space--time interaction CAR model proposed by
\citet{SchHel04}, the dynamic CAR model proposed by
Mart{\'{\i}}nez-Beneito, L{\'o}pez-Quilez and Botella-Rocamora (\citeyear{MarLopBot08}), the proper Gaussian MRF process models
of \citet{VivFer09} and the latent structure models approach
from \citet{Lawetal10}.

Our manuscript departs from this rich literature by considering the
setting where space is discrete and time is continuous. This can be
envisioned when, for instance, we have a collection of {$N_s$ functions
of time over $N_s$ regions}, but the functions are posited to be
spatially associated. That is, functions arising from neighboring
regions are believed to resemble each other. The functional data
analysis literature [Ramsay and Silverman (\citeyear{RamSil05}) and references
therein] deals almost exclusively with kernel smoothers and
roughness-penalty type (spline) models; recent discrete-space,
continuous time examples using spline-based methods include the works
by \citet{MacGus07} and Ugarte, Goicoa and Militino (\citeyear{UgaGoiMil10}).
\citet{Baletal08} consider spatially correlated
functional data modeling for point-referenced data by treating space as
continuous. A recent review by \citet{Deletal10} reveals that
spatially associated functional modeling of time has received little
attention, especially for regionally aggregated data. This is
unfortunate, especially given the data set we encounter here (see
Section~\ref{SecData} below).

As such, we propose a rich class of Bayesian space--time models based
upon a dynamic MRF that evolves continuously over time. This
accommodates spatial processes that are posited to be spatially indexed
over a geographical map with a well-defined system of neighbors. This
continuous temporal evolution sets our current article apart from the
existing literature. Rather than modeling time using simple parametric
forms, as is often done in longitudinal contexts, we employ a
stochastic process, enhancing the model's adaptability to the data.

The benefits of using a continuous-time model over a discrete-time
model here are twofold. First and foremost, investigators (or, in our
setting, public health officials) may desire understanding of the local
effects of temporal impact at a resolution finer than that at which the
data were sampled. For instance, despite collecting data monthly, there
may be interest in making inference on a particular week or even a
given day of that month. While there is a wealth of literature in this
domain, dynamic space--time models that treat time discretely can offer
statistically legitimate inference only at the level of the data.
Second, the modeling also allows us to subsequently carry out inference
on temporal gradients, that is, the rate of change of the underlying
process over time. We show how such inference can be carried out in
fully model-based fashion using exact posterior predictive
distributions for the gradients at any arbitrary time point.

The smoothness implications for the underlying process in this context
are obvious. We deploy a mean square differentiable Gaussian process
that provides a tractable gradient (or derivative) process to help us
achieve these inferential goals. Here our goal is to detect temporal
changes in the residuals that remain after accounting for important
covariates; significant changes may correspond to changes in
spatiotemporal covariates still missing from our model. While the
residuals themselves could be beneficial in detecting missing
covariates, temporal gradients can be more useful in detecting
covariates that operate on much finer scales. For example, time points
with significantly high residual gradients are likely to point toward
missing covariates whose rapid changes on a finer scale impact the
outcome. On the other hand, the residual process estimated from
discrete time models is likely to smooth over any patterns arising from
such local behavior of covariates.

The remainder of the manuscript is structured as follows. Section~\ref
{SecData} describes the data set that motivates our methodology and
which we analyze in depth. Section~\ref{SecArealTemporalProcesses}
outlines a class of dynamic MRF indexed continuously over time.
Section~\ref{SecHierarchicalModeling} provides details on the
Bayesian hierarchical models that emerge from our rich space--time
structures, while Section~\ref{SecGradientAnalysis} derives the
posterior predictive inferential procedure for the temporal gradient
process, verified via simulation in Section~\ref{SecSim}.
Section~\ref
{SecAnalysis} describes the detailed analysis of our data set, while
Section~\ref{SecDiscussion} summarizes and concludes.

\section{Data}\label{SecData}

Our
data set
consists of asthma hospitalization rates in the state of California.
According to the \citet{CAoHS03},
millions of residents of California suffer from asthma or asthma-like
symptoms. As many studies have indicated [e.g., \citet{Engetal98}], asthma rates are related to, among other things, pollution
levels and socioeconomic status (SES)---two variables that likely
induce a spatiotemporal distribution on such rates. Weather and climate
also likely play a role, as cold air can trigger asthma symptoms.

\begin{figure}

\includegraphics{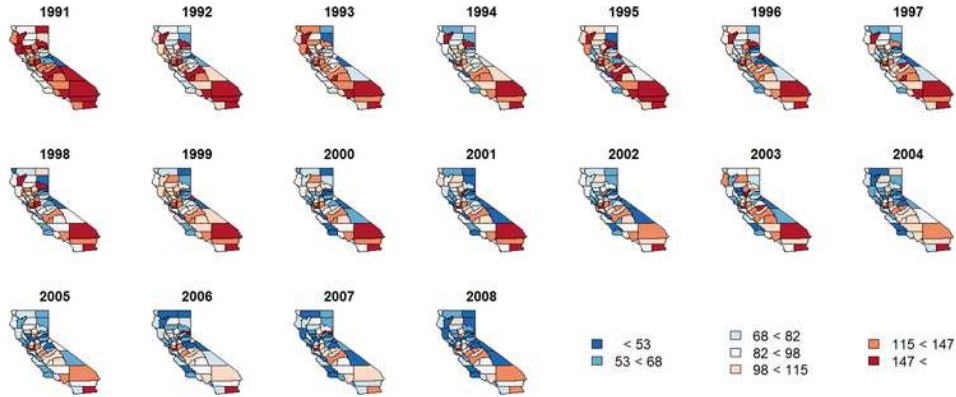}

\caption{Raw annual (summed over month) asthma hospitalization rates
per 100,000. Note: the analysis performed here was conducted on the
\emph{monthly} level; annual aggregation for illustration purposes only.}
\label{figasthmaannual}
\end{figure}

The data we will analyze were collected daily from 1991 to 2008 {from
each of the 58 counties}. We consider all hospital discharges where
asthma was the primary diagnosis, which are categorized as extrinsic
(allergic), intrinsic (nonallergic) or other. Due to confidentiality,
data for days with between one and four hospitalizations of a specific
category are missing; this affected 38\% of our observations, including
more than 50\% of those from 21 counties. To remedy this,
county-specific values for these days are imputed using a method
similar to Besag's iterated conditional modes method [\citet{Bes86}];
see the online supplement [Quick, Banerjee and Carlin (\citeyear{QuiBanCar})] for details. For our
analysis, the data are aggregated by month, for a total of 216
observations per county over the 18-year period, and then \emph{rates}
per 100,000 residents are computed; the conversion from counts to rates
for the purpose of fitting Gaussian spatiotemporal models is common {in
the literature} [see, e.g., Short, Carlin and Bushhouse (\citeyear{ShoCarBus02})].
While the vast majority of rates are less than 20 hospitalizations per
month per 100,000 people, the range of the rates extends from 0 to 90.
As can be seen in Figure~\ref{figasthmaannual}, hospitalization for
asthma demonstrates a statewide decreasing trend early in the study
period and appears to stabilize in later years. Here, we map the raw
\emph{annual} (summed over month) hospitalization rates, which have
values between 0 and 340 hospitalizations per 100,000.

We attempt to capture the effect of socioeconomic status by including
population density in our model, using data from the 2000 U.S. Census
and land area measurements from the National Association of Counties.
To account for pollution, we use data from the Air Resources Board of
the California Environmental Protection Agency which counts the number
of days in each month with average ozone levels above 0.07 ppm over 8
consecutive hours, the state standard. Because our ozone data is
compiled at the \textit{air basin} level, county-specific values are
calculated by taking the maximum value of all air basins that the
county belonged to. Generally, ozone levels are highest during the
summer months, with the highest values in southern California and the
Central Valley region, and show little variation between years.
As
hospitalization rates are higher among youth and the black population,
county-level covariates for percent under 18 and percent black are also
included. These demographic covariates both have their highest values
in southern California, though counties in the Central Valley region
also have larger black populations.

%


\section{Areally referenced temporal processes}\label{SecArealTemporalProcesses}

As mentioned above, our\break methodological contribution is a modeling
framework for areally referenced outcomes that, it can be reasonably
assumed, arise from an underlying stochastic process continuous over
time. To be specific, consider a map of a geographical region
comprising $N_s$ regions that are delineated by well-defined
boundaries, and let $Y_i(t)$ be the outcome arising from region $i$ at
time $t$. For every region $i$, we believe that $Y_i(t)$ exists, at
least conceptually, at every time point. However, the observations are
collected not continuously but at discrete time points, say, $\mathcal{
T} = \{t_1, t_2,\ldots, t_{N_t}\}$. For the time being, we will assume
that the data comes from the same set of time points in $\mathcal{ T}$
for each region. This is not necessary for the ensuing development, but
will facilitate the notation.

A spatial random effect model for our data assumes
%
\begin{eqnarray}
\label{EqBasicModel} Y_i(t) = \mu_i(t) +
Z_i(t) + \varepsilon_i(t),\qquad \varepsilon_i(t)
\stackrel{\mathrm{ind}} {\sim} N\bigl(0,\tau_i^2\bigr)
\nonumber
\\[-8pt]
\\[-8pt]
\eqntext{\mbox{for }
i=1,2,\ldots ,N_s ,}
\end{eqnarray}
%
where $\mu_i(t)$ captures large scale variation or trends, for example,
using a regression model, and $Z_i(t)$ is an underlying
areally-referenced stochastic process over time that captures
smaller-scale variations in the time scale while also accommodating
spatial associations. Each region also has its own variance component,
$\tau_i^2$, which captures residual variation not captured by the other
components.

The process $Z_i(t)$ specifies the probability distribution of
correlated space--time random effects while treating space as discrete
and time as continuous. We seek a specification that will allow
temporal processes from neighboring regions to be more alike than from
nonneighbors. As regards spatial associations, we will respect the
discreteness inherent in the aggregated outcome. Rather than model an
underlying response surface continuously over the region of interest,
we want to treat the $Z_i(t)$'s as functions of time that are smoothed
across neighbors.

The neighborhood structure arises from a discrete topology comprising a
list of neighbors for each region. This is described using an
$N_s\times N_s$ adjacency matrix $W = \{w_{ij}\}$, where $w_{ij}=0$ if
regions $i$ and $j$ are not neighbors and $w_{ij}=c\ne0$ when regions
$i$ and $j$ are neighbors, denoted by $i\sim j$. By convention, the
diagonal elements of $W$ are all zero. To account for spatial
association in the $Z_i(t)$'s, a~temporally evolving MRF for the areal
units at any arbitrary time point $t$ specifies the full conditional
distribution for $Z_i(t)$ as depending only upon the neighbors of
region $i$,
%
\begin{equation}
\label{EqMarkovRandomField} p\bigl(Z_{i}(t) | \bigl
\{Z_{j\neq i}(t)\bigr\}\bigr)\sim N \biggl(\sum_{j\sim i}
\alpha \frac
{w_{ij}}{w_{i+}}Z_{j}(t), \frac{\sigma^2}{w_{i+}} \biggr),
\end{equation}
where $w_{i+} = \sum_{j\sim i} w_{ij}$, $\sig^2 > 0$, and $\alpha$
is a
propriety parameter described below. This means that the $N_s\times1$
vector $\bZ(t) = (Z_1(t), Z_2(t),\ldots, Z_{N_s}(t))^{T}$ follows a
multivariate normal distribution with zero mean and a precision matrix
$\frac{1}{\sigma^2}(D - \alpha W)$, where $D$ is a diagonal matrix with
$w_{i+}$ as its $i$th diagonal elements. The precision matrix is
invertible as long as $\alpha\in(1/\lambda_{(1)},1/\lambda_{(n)})$,
where $\lambda_{(1)}$ (which can be shown to be negative) and $\lambda_{(n)}$
(which can be shown to be 1) are the smallest (i.e., most
negative) and largest eigenvalues of $D^{-1/2}WD^{-1/2}$, respectively,
and this yields a proper distribution for $\bZ(t)$ at each time point $t$.

The MRF in (\ref{EqMarkovRandomField}) does not allow temporal
dependence; the $\bZ(t)$'s are independently and identically
distributed as $  N (\bzero, \sigma^2(D - \alpha
W)^{-1} )$. We could allow time-varying parameters $\sigma_t^2$
and $\alpha_t$ so that $\bZ(t) \stackrel{\mathrm{ind}}{\sim} N (\bzero,
\sigma_t^2(D - \alpha_t W)^{-1} )$ for every~$t$. If time were
treated discretely, then we could envision dynamic autoregressive
priors for these time-varying parameters, or some transformations
thereof. However, there are two reasons why we do not pursue this
further. First, we do not consider time as discrete because that would
preclude inference on temporal gradients, which, as we have mentioned,
is a major objective here. Second, time-varying hyperparameters,
especially the $\alpha_t$'s, in MRF models are usually weakly
identified by the data; they permit very little prior-to-posterior
learning and often lead to over-parametrized models that impair
predictive performance over time.

Here we prefer to jointly build spatial-temporal associations into the
model using a multivariate process specification for $\bZ(t)$. A highly
flexible and computationally tractable option is to assume that $\bZ
(t)$ is a zero-centered multivariate Gaussian process, $\operatorname{GP}(\bzero,
K_Z(\cdot,\cdot))$, where the {matrix-valued covariance function [e.g.,
``\emph{cross-covariance} matrix function,'' \citet{Cre93}]} $K_Z(t,u)
= \operatorname{cov}\{\bZ(t),\bZ(u)\}$ is defined to be the
$N_s\times N_s$
matrix with $(i,j)$th entry $\operatorname{cov}\{Z_i(t),  Z_j(u)\}$ for any
$(t,u)\in\Re^{+}\times\Re^{+}$. Thus, for any two positive real
numbers $t$ and $u$, $K_Z(t,u)$ is an $N_s\times N_s$ matrix with
$(i,j)$th element given by the covariance between $Z_i(t)$ and
$Z_j(u)$. These multivariate processes are \emph{stationary} when the
{covariances} are functions of the separation between the time points,
in which case we write $K_Z(t,u)=K_Z(\Delta)$, and \emph{fully
symmetric} when $K_Z(t,u)=K_Z(|\Delta|)$, where $\Delta= t-u$. For a
detailed exposition on 
{covariance} functions,
see Chapter 7 of Banerjee, Gelfand and Sirmans (\citeyear{BanGelSir03}); \citet{GelBan10}
and \citet{GneGut10} also provide overviews for continuous settings.

To ensure valid joint distributions for process realizations, we use a
constructive approach similar to that used in \emph{linear models of
coregionalization} (LMC) and, more generally, belonging to the class of
multivariate latent process models [see Section 7.2 of Banerjee, Gelfand and Sirmans (\citeyear{BanGelSir03})].
We assume that $\bZ(t)$ arises as a (possibly temporally-varying)
linear transformation $\bZ(t)=A(t)\bv(t)$ of a simpler process $\bv(t)=
(v_1(t),v_2(t),\ldots,v_{N_s}(t))^{T}$, where the $v_{i}(t)$'s are
univariate temporal processes, independent of each other, and with unit
variances. This differs from the conventional LMC approach based on
\emph{spatial} processes, {which treats space} as continuous.
The matrix-valued covariance function 
for $\bv(t)$, say, $K_{\bv}(t,u)$, thus has a simple diagonal form and
$K_Z(t,u)=A(t)K_{\bv}(t,u)A(u)^{T}$. The dispersion matrix for $\bZ$ is
$\Sigma_{Z}=\mathcal{ A}\Sigma_\bv\mathcal{ A}^{T}$, where
$\mathcal{
A}$ is a block-diagonal matrix with $A(t_j)$'s as blocks, and $\Sigma_{\bv}$
is the dispersion matrix constructed from $K_{\bv}(t,u)$.
Constructing simple valid 
{matrix-valued covariance functions} for $\bv(t)$ automatically ensures
valid probability models for $\bZ(t)$. Also note that for $t=u$,
$K_{\bv
}(t,t)$ is the identity matrix so that $K_Z(t,t)=A(t)A(t)^{T}$ and
$A(t)$ is a square-root (e.g., obtained from the triangular Cholesky
factorization) of the matrix-valued covariance function
at time $t$.

The above framework subsumes several simpler and more intuitive
specifications. One particular specification that we pursue here
assumes that each $v_{i}(t)$ follows a stationary Gaussian Process
$\operatorname{GP}(0,\rho(\cdot,\cdot;\bphi))$, where $\rho(\cdot,\cdot; \phi
)$ is a
positive definite correlation function parametrized by $\bphi$
[e.g., \citet{Ste99}], so that $\operatorname{cov}(v_i(t), v_i(u)) = \rho
(t,u;\phi)$
for every $i=1,2,\ldots,N_s$ for all nonnegative real numbers $t$ and
$u$. Since the $v_i(t)$ are independent across $i$, $\operatorname
{cov}\{
v_i(t),v_j(u)\} = 0$ for $i\neq j$.

The matrix-valued covariance function 
for $\bZ(t)$ becomes $K_Z(t,u)= \rho(t,u;\break\bphi)A(t)A(u)^{T}$. If we
further assume that $A(t)=A$ is constant over time, then the process
$\bZ(t)$ is stationary if and only if
$\bv(t)$ is stationary. 
Further, we obtain a \emph{separable} specification, so that
$K_Z(t,u)=\rho(t,u;\bphi)AA^{T}$. Letting $A$ be some square-root
(e.g., Cholesky) of the $N_s\times N_s$ dispersion matrix $\sigma^2(D-\alpha W)^{-1}$
and $R(\bphi)$ be the $N_t\times N_t$ temporal correlation matrix
having $(i,j)$th element $\rho(t_i,t_j;\bphi)$
yields
%
\begin{eqnarray}
\label{EqSeparableCrossCov} 
K_Z(t,u)&=&
\sigma^2\rho(t,u;\bphi) (D-\alpha W)^{-1} \quad\mbox{and}
\nonumber
\\[-8pt]
\\[-8pt]
\nonumber
\Sigma_{Z}&=&R(\bphi)\otimes\sigma^2(D-\alpha
W)^{-1} .
\end{eqnarray}
%
It is straightforward to show that the marginal distribution from this
constructive approach for each $\bZ(t_i)$ is $  N
(\bzero, \sigma^2(D - \alpha W)^{-1} )$, the same marginal
distribution as the temporally independent MRF specification in (\ref
{EqMarkovRandomField}). Therefore, our constructive approach
ensures a valid space--time process, where associations in space are
modeled discretely using a MRF, and those in time through a continuous
Gaussian process.

This separable specification is easily interpretable, as it factorizes
the dispersion into a spatial association component (areal) and a
temporal component. Another significant practical advantage is its
computational feasibility. Estimating more general space--time models
usually entails matrix factorizations with $O(N_s^3N_t^3)$
computational complexity. The separable specification allows us to
reduce this complexity substantially by avoiding factorizations of
$N_sN_t\times N_sN_t$ matrices. One could design algorithms to work
with matrices whose dimension is the smaller of $N_s$ and $N_t$,
thereby accruing massive computational gains.

More general models using this approach are introduced and discussed in
the online supplement [Quick, Banerjee and Carlin (\citeyear{QuiBanCar})], but since they do not
offer anything new in terms of temporal gradients, we do not pursue
them in the remainder of this paper.
\section{Hierarchical modeling}\label{SecHierarchicalModeling}

In this section we build a hierarchical modeling framework to analyze
the data in Section~\ref{SecData} using the likelihood from our
spatial random effects model in (\ref{EqBasicModel}) and the
distributions emerging from the temporal Gaussian process discussed in
Section~\ref{SecArealTemporalProcesses}. The mean $\mu_i(t)$ in
(\ref{EqBasicModel}) is often indexed by a parameter vector $\bbeta
$, for example, a linear regression with regressors indexed by space
and time so that $\mu_i(t;\bbeta) = \bx_i(t)^{T}\bbeta$.

The posterior distributions we seek can be expressed as
%
\begin{eqnarray}
\label{EqPosteriorDistribution} p(\btheta, \bZ\given\bY) &\propto& p(\bphi)\times
\operatorname{IG}\bigl(\sigma^2\given a_{\sigma}, b_{\sigma}\bigr)
\times \Biggl(\prod_{i=1}^{M} \operatorname{IG}\bigl(
\tau_i^2 | a_{\tau},b_{\tau}\bigr)
\Biggr) \times N(\bbeta\given\mu_{\beta
},\Sigma_{\beta}) \nonumber\\
&&{}\times
\operatorname{Beta}(\alpha\given a_{\alpha},b_{\alpha})
\nonumber
\\
&&{} \times N\bigl(\bZ\given\bzero, R(\phi) \otimes\sigma^2 (D-\alpha
W)^{-1}\bigr)\\
&&{} \times\prod_{j=1}^{N_t}
\prod_{i=1}^{N_s} N\bigl(Y_i(t_j)
\given \bx_i(t_j)^{T}\bbeta+
Z_i(t_j), \tau_i^2\bigr) ,\nonumber
\end{eqnarray}
where $\btheta= \{\bphi,\alpha,\sigma^2, \bbeta,\tau_1^2,\tau_2^2,\ldots,\tau_{N_s}^2\}$ and $\bY$ is the vector of observed
outcomes defined analogous to $\bZ$. The parametrizations for the
standard densities are as in \citet{CarLou09}. We assume all
the other hyperparameters in (\ref{EqPosteriorDistribution}) are known.

Recall the {separable matrix-valued covariance function}
in (\ref{EqSeparableCrossCov}). The correlation function $\rho
(\cdot
;\bphi)$ determines process smoothness and we choose it to be a fully
symmetric Mat\'ern correlation function given by
%
\begin{equation}\qquad
\rho(t,u;\bphi) = \rho(\Delta;\bphi) = \frac{1}{\Gamma(\phi_2)2^{\phi
_2-1}}\bigl (2\sqrt{
\phi_2}|\Delta|\phi_1 \bigr)^{\phi_2} \mathcal{
K}_{\phi_2} \bigl(2\sqrt{\phi_2}|\Delta|\phi_1 \bigr),
\end{equation}
where $\bphi= \{\phi_1,\phi_2\}$, $\Delta= t-u$, $\Gamma(\cdot)$ is
the Gamma function, $\mathcal{ K}_{\phi_2}(\cdot)$ is the modified
Bessel function of the second kind, and $\phi_1$ and $\phi_2$ are
nonnegative parameters representing rate of decay in temporal
association and smoothness of the underlying process, respectively.

We use Markov chain Monte Carlo (MCMC) to evaluate the joint posterior
in~(\ref{EqPosteriorDistribution}), using Metropolis steps for
updating $\bphi$ and Gibbs steps for all other parameters, details of
which are shown in the supplemental article [Quick, Banerjee and Carlin (\citeyear{QuiBanCar})].
Sampling-based Bayesian inference seamlessly delivers inference on the
residual spatial effects. Specifically, if $t_0$ is an arbitrary
unobserved time point, then, for any region $i$, we sample from the
posterior predictive distribution $p(Z_i(t_0)\given\bY) = \int
p(Z_i(t_0)\given\bZ,\btheta)  p(\btheta,\break\bZ\given\bY) \,d\btheta
\,d\bZ
$. This is achieved using \emph{composition sampling}: for each sampled
value of $\{\btheta, \bZ\}$, we draw $Z_i(t_0)$, one for one, from
$p(Z_i(t_0)\given\bZ,\btheta)$, which is Gaussian. Also, our sampler
easily adapts to situations where $Y_i(t)$ is missing (or not
monitored) for some of the time points in region $i$. We simply treat
such variables as missing values and update them, from their associated
full conditional distributions, which of course are $N(\bx_i(t)^T\bbeta
+ Z_i(t), \tau^2_i)$. We assume that all predictors in $\bx_i(t)$ will
be available in the space--time data matrix, so this temporal
interpolation step for missing outcomes is straightforward and inexpensive.

Model checking is facilitated by simulating \emph{independent}
replicates for each observed outcome: for each region $i$ and observed
time point $t_j$, we sample from $p(Y_{\mathrm{rep},i}(t_j)\given\bY) = \int
N(Y_{\mathrm{rep},i}(t_j)\given\bx_i(t_j)^T\bbeta+ Z_i(t_j), \tau_i^2)
p(\bbeta, Z_i(t_j),\break \tau_i^2\given\bY) \,d\bbeta \,dZ_i(t_j) \,d\tau_i^2$,
where
$p(\bbeta, Z_i(t_j),\tau_i^2\given\bY)$ is the marginal posterior
distribution of the unknowns in the likelihood. Sampling from the
posterior predictive distribution is straightforward, again, using
composition sampling.
\section{Gradient analysis}\label{SecGradientAnalysis}

Our primary goal is to carry out statistical inference on temporal
gradients with data arising from a temporal process indexed discretely
over space. We will do so using the notions of smoothness of a Gaussian
process and its derivative. \citet{Adl09}, \citet{Maretal96} and
\citet{BanGel03} discuss derivatives (more generally, linear
functionals) of Gaussian processes, while Banerjee, Gelfand and Sirmans
(\citeyear{BanGelSir03}) lay out an inferential framework for directional gradients on a
spatial surface. Most of the existing work on derivatives of stochastic
processes deal either with purely temporal or purely spatial processes
[see, e.g., \citet{Ban10}]. Here, we consider gradients for a
temporal process indexed discretely over space.

Assume that $ \{Z_i (t )\dvtx t\in{\Re^{+}} \}$ is a
stationary random process for each region $i$.\footnote{Stationarity is
not required. We only use it to ensure smoothness of realizations and
to simplify forms for the induced 
covariance function.} The process is $L_{2}$ (or mean
square) continuous at $t_{0}$ if {$\mathop{\lim}_{t\rightarrow
t_{0}}E\llvert Z_i (t )-Z_i (t_{0} ) \rrvert^{2}=0$}.
The notion of a mean square differentiable process can be formalized
using the analogous definition of total
differentiability of a function in a nonstochastic setting [see, e.g.,
\citet{BanGel03}]: $Z_i (t )$ is mean square
differentiable at $t_{0}$ if it admits a first order linear expansion
for any scalar $h$,
%
\begin{equation}
Z_i (t_{0}+h ) = Z_i (t_{0}
)+hZ_i'(t)+o (h ) \label{FirstOrderLinearity}
\end{equation}
in the $L_{2}$ sense as $h\rightarrow0$, where we say that
$ \frac{d}{dt}Z_i(t) = Z_i' (t_{0} )$ is the
\emph
{gradient} or \emph{derivative} process derived from the \emph{parent}
process $Z_i(t)$. In other
words, we require
%
{\renewcommand{\theequation}{\arabic{equation}$^\prime$}
\setcounter{equation}{5}
\begin{equation}
\label{msderiv} \mathop{\lim}_{h\rightarrow0}E \biggl(\frac{Z_i (t_{0}+h )
-Z_i (t_{0} )}{h}-Z_i'(t_0)
\biggr)^{2}=0. 
\end{equation}}
\hspace*{-3pt}Equations~(\ref{FirstOrderLinearity}) and (\ref{msderiv}) ensure that
mean square differentiable processes are mean square continuous.

For a univariate stationary process, smoothness in the mean square
sense is determined by its covariance or correlation function.
A stationary multivariate process $\bZ(t)$ with {matrix-valued
covariance function} 
$K_Z(\Delta)$ will admit a well-defined gradient process
$\bZ'(t) = (Z_1'(t),\ldots,Z_{N_s}'(t))^{T}$
if and only if $K_Z''(0)$ exists, where $K_Z''(0)$ is the element-wise
second-derivative of $K_Z(\Delta)$ evaluated at $\Delta=0$.\vadjust{\goodbreak}

A Gaussian process with a Mat\'ern correlation function has sample
paths that are $\lceil\phi_2-1 \rceil$ times differentiable. As
$\phi_2\to\infty$, the Mat\'ern correlation function converges to the
squared exponential (or the so-called Gaussian) correlation function,
which is infinitely differentiable and leads to acute oversmoothing.
When $\phi_2=0.5$, the Mat\'ern correlation function is identical to
the exponential correlation function [see, e.g., \citet{Ste99}]. To
ensure that the underlying process is differentiable so that the
gradient process exists, we need to restrict $\phi_2 > 1$. However,
letting $\phi_2 > 2$ usually leads to oversmoothing, as the data can
rarely distinguish among values of the smoothness parameter greater
than $2$. Hence, we restrict $\phi_2 \in(1,2]$. We could either assign
a prior on this support or {simply fix $\phi_2$} somewhere in this
interval. Since it is difficult to elicit informative priors for the
smoothness parameter, we would most likely end up with a uniform prior.
In our experience, not only does this deliver only modest posterior
learning and lead to an increase in computing (both in terms of MCMC
convergence and estimating the resulting correlation function and its
derivative), but the substantive inference is almost indistinguishable
from what is obtained by fixing $\phi_2$.

As such, in our subsequent analysis we fix $\phi_2 = 3/2$, which has
the side benefit of yielding the closed form expression $\rho(\Delta
;\phi_1) = (1+\phi_1 |\Delta|)\times\break\exp(-\phi_1 |\Delta|)$. The first
and second order derivatives for the {matrix-valued covariance
function} 
in~(\ref{EqSeparableCrossCov}) can now be obtained explicitly as
%
\begin{eqnarray}
\label{EqDerivativesCrossCovarianceFunction} K_{Z}'(
\Delta) &= &-\sigma^2 \phi_1^2 \Delta\exp\bigl(-
\phi_1 |\Delta|\bigr) (D-\alpha W)^{-1}\quad\mbox{and}
\nonumber
\\[-8pt]
\\[-8pt]
\nonumber
-K_{Z}''(0) &=& \sigma^2
\phi_1^2 (D-\alpha W)^{-1} .
\end{eqnarray}

Turning to inference for gradients, we seek the joint posterior
predictive distribution,
%
\begin{eqnarray}
\label{EqPosteriorPredictiveGradients} p\bigl(\bZ'(t_0)
\given\bY\bigr) &=& \int p\bigl(\bZ'(t_0) \given\bY, \bZ,
\btheta\bigr) p(\bZ\given\btheta, \bY) p(\btheta\given\bY) \,d\btheta \,d\bZ
\nonumber
\\[-8pt]
\\[-8pt]
\nonumber
&=& \int p\bigl(\bZ'(t_0) \given\bZ, \btheta\bigr) p(
\bZ | \btheta, \bY ) p(\btheta | \bY) \,d\btheta \,d\bZ ,
\end{eqnarray}
where the second equality follows from the fact that the gradient
process is derived entirely from the parent process and so $p(\bZ'(t_0)
\given\bY, \bZ, \btheta)$ does not depend on $\bY$.

We evaluate (\ref{EqPosteriorPredictiveGradients}) using
composition sampling. Here, we first obtain $\btheta^{(1)},\btheta^{(2)},\break\ldots, \btheta^{(M)} \sim p(\btheta |  \bY)$ and $\bZ^{(j)}
\sim p(\bZ |  \btheta^{(j)},\bY), j = 1,2,\ldots,M$, where $M$ is the
number of (post-burn-in) posterior samples. Next, for each $j$ we draw
$\bZ^{(j)} \sim p(\bZ |  \btheta^{(j)}, \bY)$, and finally $\bZ'(t_0)^{(j)} \sim p(\bZ'(t_0)\given\bZ^{(j)}, \btheta^{(j)})$. The
conditional distribution for the gradient can be seen to be
multivariate normal
with mean and variance-covariance matrix given by
\begin{eqnarray*}
\bmu_{Z'\given Z, \theta} &=& \operatorname{cov}\bigl(\bZ'(t_0),
\bZ\bigr) \operatorname{var}(\bZ )^{-1} \bZ= -\bigl(K_{Z}'
\bigr)^{T} \Sigma_{Z}^{-1} \bZ\quad\mbox{and}
\\
 \Sigma_{Z'\given Z, \theta} &=& -K_{Z}''(0)
- \bigl(K_{Z}'\bigr)^{T}\Sigma_{Z}^{-1}
\bigl(K_{Z}'\bigr) ,
\end{eqnarray*}
where $\Sigma_{Z}^{-1} = \frac{1}{\sigma^2} R(\phi)^{-1} \otimes
(D-\alpha W)$ and $(K_{Z}')^{T}$ is an $N_s \times N_s N_t$ block
matrix whose $j$th block is given by the $N_s\times N_s$ matrix
$K_Z'(\Delta_{0j})$, with $\Delta_{0j} = t_j - t_0$.
Note that $\Sigma_{Z'\given Z, \theta}$ is an $N_s N_t \times N_s N_t$
matrix, but we can use the properties of the MRF to only invert
$N_t\times N_t$ matrices. 

\section{Simulation studies}\label{SecSim}
{To validate our model's ability to correctly estimate both our model
parameters and the underlying temporal gradients, we have constructed
two separate simulation studies using the $N_s = 58$ counties of
California as our spatial grid and $N_t = 50$ observations per county,
where $\mathcal{ T} = \{1, 2,\ldots, 50\}$. Each simulation study
consists of 100 data sets comprised of 2900 observations generated
from~(\ref{EqBasicModel}), where $\mu_i(t) = \bx_i(t)^{T} \bbeta$,
using the same parameter values, and our results are based on 5000
MCMC samples after a burn-in period of 5000 iterations.

In an effort to obtain simulated outcomes comparable to those from our
real data, our first simulation study uses an intercept and the four
covariates described in Section~\ref{SecData}, and we set the
$5\times
1$ vector, $\bbeta$, as the least squares estimates from our real data.
We also set $\phi= 1$, $\alpha=0.90$, and $\sig^2 = 18$, which are
then used to generate true values for $\bZ$, while our $\tau_i^2$ are
drawn from an inverse Gamma distribution centered at 1 with modest
variance. For each of the 100 simulated data sets, we constructed $95\%
$ Bayesian credible intervals for each parameter and recorded the
number of times they included their true values (i.e., their
``frequentist coverage''). We found this coverage to be between 93--97\% for
the 5 $\beta$'s, about $87\%$ for the random effect variance
$\sigma^2$ and around $90\%$ on the average for the $58$ $\tau_i^2$'s,
with the majority of them having $95\%$ coverage. Coverage was poor for
$\tau_i^2 < 0.15$; in situations where small variances are to be
expected, this issue could be avoided or alleviated by rescaling the
data or specifying a prior with a larger mass near 0, respectively. The
spatiotemporal random effects, $\bZ$, also enjoyed satisfactory
coverage; the average coverage over the 2900 space--time random effects
was around $95.5\%$. By contrast, the coverage for the propriety
parameter, $\alpha$, and the spatial range parameter, $\phi$, reveal
biases, with coverages less than $50\%$. This is not entirely
unexpected, as spatial and temporal range parameters of this type are
known to be weakly identified by the data [e.g., \citet{Zha04}].
Furthermore, the biases for $\phi$ and $\alpha$ are not substantial,
with their posterior medians only 8\% above and 5\% below their true
values, respectively. In an effort to verify the robustness of our
model to these biases, we repeated the simulation with both $\phi$ and
$\alpha$ fixed at their true values and were able to reproduce our results.}

\begin{figure}

\includegraphics{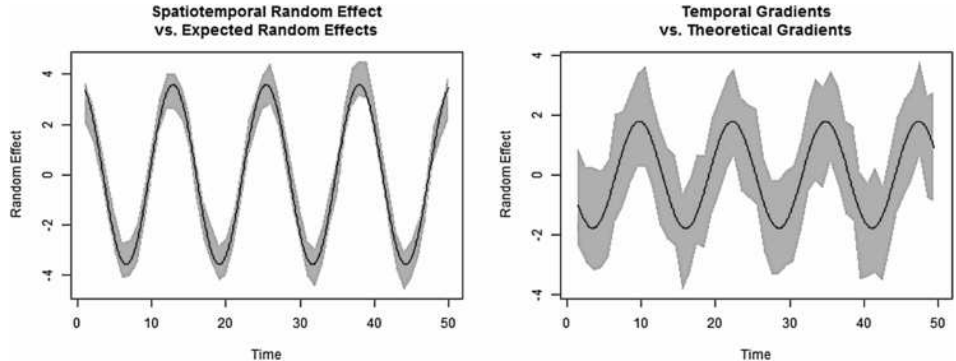}

\caption{Spatiotemporal random effects and temporal gradients for a
region based on one data set from the second simulation study. Solid
black lines denote true sinusoidal curves based on the model in
equation~(\protect\ref{eqsimsetup}), while gray bands represent
95\% credible
intervals.}
\label{figsim}
\end{figure}

Having demonstrated the ability of our model to correctly estimate
model parameters, the focus of our second simulation study is to
validate the theory of our temporal gradient processes. To do this, we assumed
%
\begin{equation}
Y_i(t_j) \ind N \biggl(5 + x_{i1}*
\operatorname{sin} \biggl(\frac
{t_j}{2} \biggr) + x_{i2}*
\operatorname{cos} \biggl(\frac{t_j}{2} \biggr), \tau_i^2
\biggr), \label{eqsimsetup}
\end{equation}
where $x_{i1}$ is the $i$th county's percent black and $x_{i2}$ is the
$i$th county's ozone level from April 1991, as described in
Section~\ref
{SecData}; this was done in order to induce spatial clustering. {As
there was no evidence of an association between the coverage of the
random effects, $\bZ$, and the region-specific variance parameters,
values of $\tau_i^2$ were generated from a $\operatorname{Uniform}(0.5,2.0)$
distribution in order to avoid the extreme values of the inverse Gamma
and focus our attention on the random effects themselves.} After
generating 100 data sets based on these parameters, we then modeled the
data using only an intercept, leaving the spatiotemporal random effects
to capture the sinusoidal curve, and conducted the gradient analysis at
the midpoints of each time interval. Figure~\ref{figsim} displays the
true spatiotemporal random effects and temporal gradients for a
particular region, along with their 95\% CI estimated from one of the
100 data sets. As can be seen, our Gaussian process model accurately
estimates both the random effects and the temporal gradients. Across
all 100 data sets, 98.3\% of the the theoretical gradients derived
using elementary calculus were covered by their respective 95\% CI,
confirming the validity of the gradient theory derived in Section~\ref
{SecGradientAnalysis}.

\section{Data analysis}\label{SecAnalysis}
As first mentioned in Section~\ref{SecData}, our data set is
comprised of monthly asthma hospitalization rates in the counties of
California over an 18-year period. As such, $N_t = 12 \cdot 18 = 216$, and
we will again use $t_j = j = 1, 2, \ldots, N_t$. The covariates in this
model include population density, ozone level, the percent of the
county under 18 and percent black. Population-based covariates are
calculated for each county using the 2000 U.S. Census, thus, they do
not vary temporally. However, the covariate for ozone level is
aggregated at the air basin level and varies monthly, though show
little variation annually. In order to accommodate seasonality in the
data, monthly fixed effects are included, using January as a baseline.
Thus, {after accounting for the monthly fixed effects and the four
covariates of interest}, $\bx_i(t)$ is a $16 \times1$ vector.

To justify the use of the model we've described, we compare it to three
alternative models using the DIC criterion [\citet{Spietal02}] and a predictive model choice criterion using strictly proper
scoring rules proposed by Gneiting and Raftery [(\citeyear{GneRaf07}) equation~(27)].
Following Czado, Gneiting and Held (\citeyear{CzaGneHel09}), we refer to this as the
Dawid--Sebastiani (D--S) score [\citet{DawSeb99}]. These
models are all still of the form
%
\begin{eqnarray}
Y_i(t) = \bx_i(t)'\bbeta+
Z_i(t) + \varepsilon_i(t),\qquad \varepsilon_i(t)
\stackrel{\mathrm{ind}} {\sim} N\bigl(0,\tau_i^2\bigr)
\nonumber
\\[-8pt]
\\[-8pt]
\eqntext{\mbox{for }
i=1,2,\ldots ,N_s ,}
\end{eqnarray}
but with different $Z_i(t)$. Our first model is a simple linear
regression model which ignores both the spatial and the temporal
autocorrelation, that is, $Z_i(t) = 0 \ \forall  i,t$. The second
model allows for a random intercept and random temporal slope, but
ignores the spatial nature of the data, that is, here $Z_i(t) = \alpha_{0i} + \alpha_{1i} t$, where $\alpha_{ki}   \iid  N(0,\sig_k^2)$,
for $k=0, 1$. In this model, to preserve model identifiability, we must
remove the global intercept from our design matrix, $\bx_i(t)$. Our
third model builds upon the second, but introduces spatial
autocorrelation by letting $\balpha_{k} = (\alpha_{k1}, \ldots,
\alpha_{kN_s})' \sim \operatorname{CAR}(\sig_k^2), k=0, 1$. The results of the model
comparison can be seen in Table~\ref{tabDIC}, which indicates that our
Gaussian process model has the {lowest DIC value and D--S score}, and is
thus the preferred model and the only one we consider henceforth. The
surprisingly large $p_D$ for the areally referenced Gaussian process
model arises due to the very large size of the data set (58 counties
$\times$ 216 time points).\looseness=-1

\begin{table}
\caption{Comparisons between our areally referenced Gaussian process
model and the three alternatives. $p_D$ is a measure of model
complexity, as it represents the effective number of parameters.
Smaller values of DIC and Dawid--Sebastiani (D--S) scores indicate a
better trade-off between in-sample model fit and model complexity}
\label{tabDIC}
\begin{tabular*}{\textwidth}{@{\extracolsep{\fill}}lccc@{}}
\hline
& \multicolumn{1}{c}{$\bolds{p_D}$} & \multicolumn{1}{c}{\textbf{DIC}\tabnoteref{t1}} & \multicolumn{1}{c@{}}{\textbf{D--S}\tabnoteref{t1}}\\
\hline
Simple linear regression & \phantom{00}79 & 9894 & 16,166\\
Random intercept and slope & \phantom{0}165 & 4347 & 10,403\\
CAR model & \phantom{0}117 & 7302 & 13,436\\
Areally referenced Gaussian process & 5256 & \phantom{000}0 & \phantom{0000.}0\\
\hline
\end{tabular*}
\tabnotetext[\textbf{*}]{t1}{Both DIC and D--S shown are standardized relative to our areally
referenced Gaussian Process model.}
\end{table}

\begin{table}
\caption{Parameter estimates for asthma hospitalization data, where
estimates for $\bar{\tau}_{\cdot}^2$ represent the median (95\% CI) of
the $\tau_i^2, i = 1,\ldots,N_s=58$}
\label{tabasthma}
\begin{tabular*}{\textwidth}{@{\extracolsep{\fill}}lclc@{}}
\hline
\textbf{Parameter} & \textbf{Median (95\% CI)} & \textbf{Parameter} & \multicolumn{1}{c@{}}{\textbf{Median (95\% CI)}} \\
\hline
$\beta_{0}$ (Intercept) & 9.17 (8.93, 9.42)\phantom{0.} & $\beta_{10}$ (July) &
$-3.78$ ($-4.21$, $-3.37$) \\
$\beta_{1}$ (Pop Den) & 0.60 (0.49, 0.70)\phantom{0.} & $\beta_{11}$ (August) &
$-3.58$ ($-4.02$, $-3.13$) \\
$\beta_{2}$ (Ozone) & $-0.18$ ($-0.28$, $-0.08$) & $\beta_{12}$ (September) &
$-1.96$ ($-2.37$, $-1.54$) \\
$\beta_{3}$ (\% Black) & 1.24 (1.15, 1.34)\phantom{0.} & $\beta_{13}$ (October)
& $-1.36$ ($-1.73$, $-1.00$) \\
$\beta_{4}$ (\% Under 18) & 1.12 (1.01, 1.24)\phantom{0.} & $\beta_{14}$ (November) &
$-0.71$ ($-1.02$, $-0.42$) \\
$\beta_{5}$ (February) & $-0.25$ ($-0.46$, $-0.04$) & $\beta_{15}$ (December)
& 0.63 (0.41, 0.86)\phantom{0.} \\
$\beta_{6}$ (March) & $-0.21$ ($-0.48$, 0.07)\phantom{0.} & $\phi$ & 0.90 (0.84, 0.97)\phantom{0.}
\\
$\beta_{7}$ (April) & $-1.47$ ($-1.81$, $-1.12$) & $\alpha$ & 0.77 (0.71,
0.80)\phantom{0.} \\
$\beta_{8}$ (May) & $-1.17$ ($-1.53$, $-0.8$)\phantom{0} & $\sig^2$ & 21.52 (20.18,
23.06)\phantom{.} \\
$\beta_{9}$ (June) & $-2.79$ ($-3.21$, $-2.4$)\phantom{0} & $\bar{\tau}_{\cdot}^2$ &
\phantom{.}3.32 (0.18, 213.16) \\
\hline
\end{tabular*}
\end{table}

The estimates for our model parameters can be seen in Table~\ref
{tabasthma}. The coefficients for the monthly covariates indicate
decreased hospitalization rates in the summer\vadjust{\goodbreak} months, a trend which is
consistent with previous findings. The coefficients for population
density, percent under 18 and percent black are all significantly
positive, also as expected. The coefficient for ozone level is
significantly negative, however, which is surprising but consistent
with the patterns in the monthly trends for both hospitalization rates
and ozone levels. This result may also be confounded by the absence of
other climate-related factors and the sensitivity of asthma admissions
to acute weather effects.\looseness=-1

There is a large range of values for the county-specific residual
variance parameters, $\tau_i^2$. Perhaps not surprisingly, the
magnitude of these terms seems to be negatively correlated with the
population of the given counties, demonstrating the effect a
(relatively) small denominator can have when computing and modeling
rates. The strong spatial story seen in the maps is reflected by the
size of $\sig^2$ compared to the majority of the $\tau_i^2$. There is
also relatively strong temporal correlation, with $\phi= 0.9$
corresponding to $\rho(t_i,t_j; \phi) \ge0.4$ for $|t_j - t_i|$ less
than 2 months.

\begin{figure}

\includegraphics{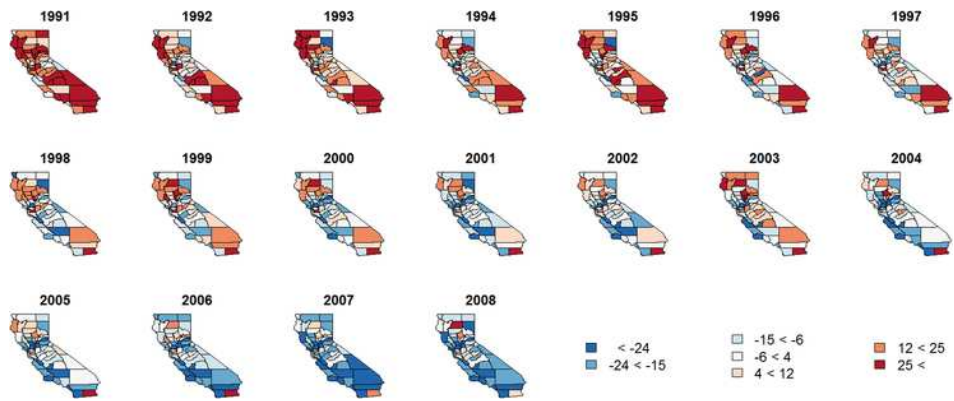}

\caption{Spatial random effects for asthma hospitalization data, by year.}
\label{figZmapannual}
\end{figure}

Maps of the yearly (averaged across month) spatiotemporal random
effects can be seen in Figure~\ref{figZmapannual}. 
Since here we are dealing with the \emph{residual} curve after
accounting for a number of mostly nontime-varying covariates, it comes
as no surprise that the spatiotemporal random effects capture most of
the variability in the model, including the striking decrease in yearly
hospitalization rates over the study period. It also appears that our
model is providing a better fit to the data in the years surrounding
2000, perhaps indicating that we could improve our fit by allowing our
demographic covariates to vary temporally. Our model also appears to be
performing well in the central counties, where asthma hospitalization
rates remained relatively stable for much of the study period.

\begin{figure}

\includegraphics{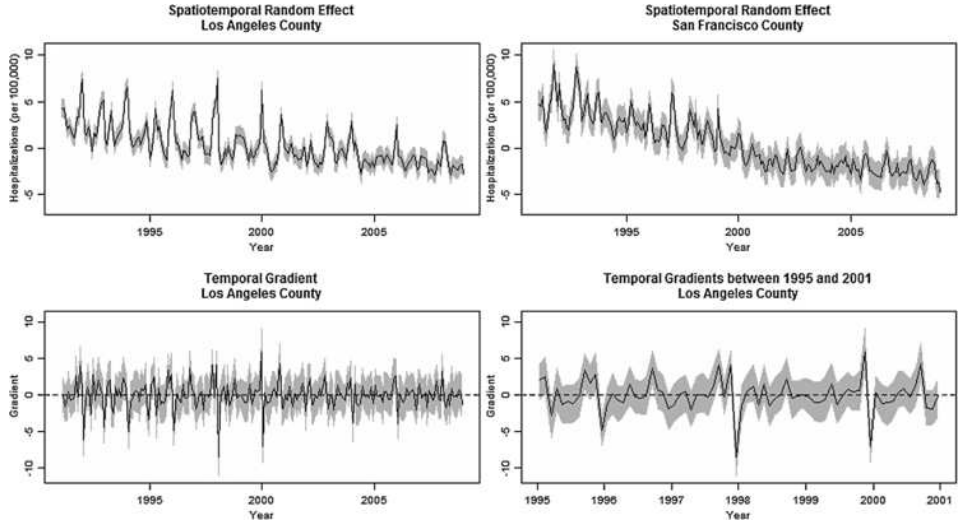}

\caption{Comparison between the spatiotemporal random effects in Los
Angeles and San Francisco Counties, and an investigation of temporal
gradients in Los Angeles County. Point estimates in black and
corresponding 95\% CI bands in gray. Figures in the top panel
illustrate the differences in the temporal trends of the random effects
between the two counties. The bottom-left figure displays the temporal
gradients computed between months in Los Angeles County, and the
bottom-right figure displays the subset of the gradients which are
further described in the text.}
\label{figLAvsSF}
\end{figure}

In the top panel of Figure~\ref{figLAvsSF}, we compare the monthly
temporal profiles of the random effects for Los Angeles and San
Francisco Counties. For Los Angeles County, the spatiotemporal random
effects (top-left panel) decrease at a consistent, moderate rate
throughout the length of the study with several large spikes prior to
2000. In contrast, San Francisco County's random effects (top-right)
have fewer and less dramatic spikes. In addition, San Francisco County
appears to have had a changepoint in its spatiotemporal random effects
around 2000, where they transition from a fairly steady decline to a
period of lower variability and very little mean change. Further
investigation may reveal a corresponding change in social,
environmental or health care reimbursement policy. The bottom-left
panel shows the temporal trend of the gradients in Los Angeles County,
which reveal the large degree of variability in the random effects. In
fact, as more clearly shown in the bottom-right panel of Figure~\ref
{figLAvsSF}, the September to October gradient was significantly
positive five times between 1995 and 2001, and three times during this
period (1995, 1997 and 1999) the November to December gradients were
significantly positive, but were immediately followed by significantly
\emph{negative} gradients from December to January, a pattern that is
seen throughout the region.


%

\begin{figure}[b]

\includegraphics{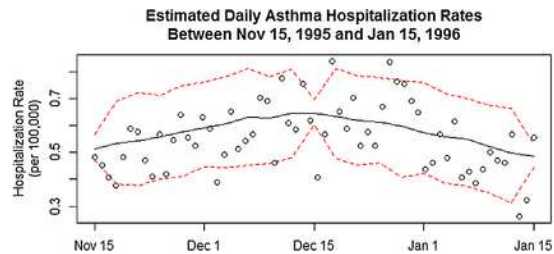}

\caption{Posterior predicted curves (and 95\% credible bounds) for the
daily asthma hospitalization rates in Los Angeles County between
November 15, 1995 to January 15, 1996. This county and interval was
selected due the presence of a significantly positive gradient between
November and December and a significantly negative gradient between
December and January. The true hospitalizations are also shown for
comparison purposes, though the model was fit using only the monthly
aggregates.}
\label{figLApred}
\end{figure}

A strength of using a continuous-time model for these data is that it
seamlessly permits prediction at a finer resolution than that of the
observed data. Upon seeing the significant gradients in Los Angeles
County in November and December of 1995, public health officials may
ask for a more detailed report than a monthly aggregation can provide.
If a discrete-time model were used, researchers would be required to
refit the model, pre-specifying at which unobserved time points to
conduct inference; however, with this model, we can use the posterior
predictive distribution to interpolate values at any time. As a
demonstration of this, Figure~\ref{figLApred} displays the predicted
daily values (solid line) and 95\% CI bands (dashed lines) every 3 days
during the period November 15, 1995 to January 15, 1996, plotted
against the true observed rates (open circles).\vadjust{\goodbreak} Despite substantial
noise in the data and modeling based solely on the aggregate rates for
each month (and assigning that value to the temporal midpoint of each
month), our predictions and 95\% CI bands perform reasonably well.

\begin{figure}

\includegraphics{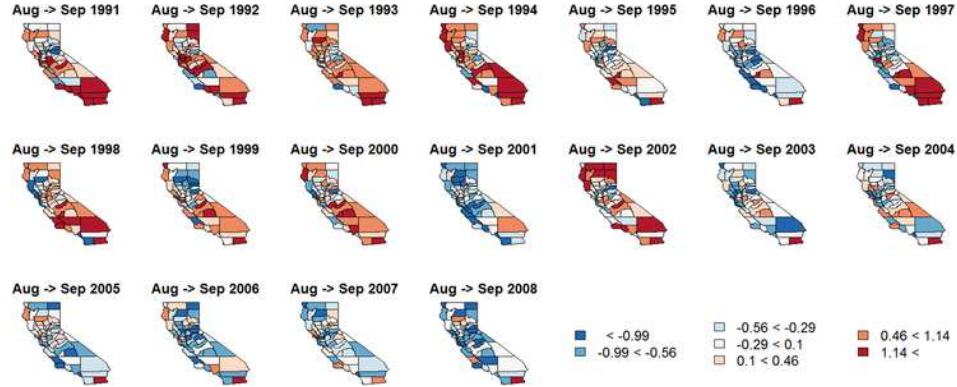}

\caption{Temporal gradients for transition from August to September
over time.}
\label{figaug2sep}
\end{figure}

As our data are aggregated monthly, we felt it was also important to
investigate the gradients on a month-to-month basis over the course of
the study. For instance, Figure~\ref{figaug2sep} reveals the gradients
between August and September decrease substantially statewide over the
course of the study. Coupling this with the information in Table~\ref
{tabasthma}, which indicates that hospitalization rates in September
are $\beta_{12} - \beta_{11} = 1.62$ per 100,000 higher than those in
August, suggests that the difference in asthma hospitalization rates
between August and September has decreased nearly 60\%, going from roughly
2.31 at the beginning of the period to just 0.97 by the end. An
investigation of the raw hospitalization rates shows a similar trend,
but this is to be expected since most of the spatiotemporal variability
in the model is accounted for by the random effects. A similar, though
not as striking, phenomenon occurs between March and April, where the
gradients are increasing. As these two pairs of months lie on the
transition between the warmer months and the cooler months, this result
would seem to suggest that the effect of seasonality has moderated over
the length of the study.\looseness=1

One limitation of this analysis is that the data records asthma
\textit{hospitalizations}, not overall \textit{prevalence}. This is an important
distinction, as factors that trigger symptoms of asthma may not be the
same as or have the same impact on asthma hospitalizations. For
instance, residents of regions with high risk environments may be
better educated about and/or prepared for managing their symptoms,
which could lead to a relative decrease in asthma hospitalization
rates. Another limitation is that, due to the aggregation of our data,
we have an inconvenient interpretation of the daily estimates in
Figure~\ref{figLApred}. A more accurate interpretation of these values
is that they are the average daily rates for the one-month interval
centered at a particular day. More generally, the interpretation of
predicted values at any time point is determined by the aggregation of
the data, but this is certainly not unique to this model.

\section{Summary and conclusions}\label{SecDiscussion}
In this paper we have provided an overview of parent and gradient
processes, building on previous work in spatiotemporal Gaussian process
modeling. We then described our modeling framework and methodology that
allows for inference on temporal gradients. An implementation of this
work was outlined in Section~\ref{SecHierarchicalModeling}, and its
theory was verified via simulation.
Its use was then illustrated on 
a real data set in Section~\ref{SecAnalysis}, where our results
showed real insight can be gained from an assessment of temporal
gradients in the residual Gaussian process, indicating overall trends
as well as motivating a search for temporally interesting covariates
still missing from our model (say, one that changes abruptly in San
Francisco County around 2000).

We believe there are two primary points of discussion regarding this
work, the first of which is the use of modeling time as continuous. If
inference is desired at the resolution of the data only, then several
of the discrete-time models in the literature would be appropriate; in
Appendix D of the online supplement to this article, we compare our
methods to one such model. Oftentimes, however, this is not the case,
as investigators and administrators may seek estimates of the temporal
effects on a finer scale. In our example, public health officials may
be interested in the daily effects of asthma, which can be correlated
with effects of daily variation of temperature and a variety of
atmospheric pollutants. A practical issue here is that hospitalization
data are often more cleanly available as monthly aggregates (say, due
to patient confidentiality issues, like those described in Section~\ref
{SecData}) and, even when the daily data are available, they tend to
be both massive and very likely to have many missing values. Analyzing
such data using discrete-time models would require methods for handling
temporal misalignment, while our temporal process-based methods can
handle such inference in a posterior predictive fashion. Furthermore,
treating time as continuous permits inference on temporal gradients,
which we feel can be an important tool for better understanding complex
space--time data sets. In some sense, our modeling framework can be
looked upon as generalizing the work of \citet{VivFer09} with
a stochastic temporal process and deriving a tractable inferential
framework for infinitesimal rates of change for that
process.\looseness=1

A second important point of discussion is the importance of
significance with respect to these temporal gradients. We believe it
depends on the problem being modeled.
While we have accounted for monthly differences in our design matrix,
the $Z_i(t)$ here may simply be capturing the remaining cyclical trend,
and this is why we felt it was more beneficial to focus on a
side-by-side comparison of two of California's most populous counties,
which motivated a further investigation of Los Angeles County, and the
trends of the twelve month-to-month comparisons rather than solely on
whether a specific gradient for a particular county was significant. In
situations where it's reasonable to assume two time points are
comparable, investigating significant temporal gradients can indicate
periods of important changes in the data, which may be caused by rapid
changes in missing covariates. We also point out that the methodology
for gradients outlined here can be applied to more general spatial
functional data analysis contexts and will be especially useful for
estimating gradients from high-resolution samples of the
function.\looseness=1

Regarding the specific application of this methodology in this paper,
it bears mentioning that modeling our data as \emph{rates} is not the
only option. Often, the counts themselves are modeled directly using a
log-linear model, with a Poisson distributional assumption justified as
a rare-events approximation to the binomial. In this setting, however,
we would no longer be able to rely on the closed form Gibbs Sampler for
updating our random effects, instead requiring Metropolis updates and a
substantial increase in computational burden. Another option is to use
a Freeman--Tukey transformation of the rates {and a single error
variance parameter, $\tau^2$, which is scaled by the county's
population}, as shown in \citet{FreTuk50} and \citet{CreCha89},
with the goal of justifying the assumption of normality. Given
the population sizes we're dealing with, we believe the assumption of
normality of our observed rates can be justified as a normal
approximation to the binomial. Furthermore, an analysis of the
transformed data results in nearly identical substantive findings.
However, there is a drawback: by modeling transformed values instead of
the rates themselves, we lose the interpretability of the scale for not
only our regression parameters, but also the temporal gradients. In our
experience, a common question among public health practitioners is,
``What does this mean?'' As such, we feel that having results which are
straightforward to interpret is of the utmost importance and, thus, we
chose to model the untransformed rates. Incidentally, we also
considered modeling the untransformed rates using a model with a single
error variance parameter (scaled by population). Sadly, the simplicity
of this model failed to outweigh its loss of flexibility and, in any
case, this model would not be generalizable to nonrate data.

One weakness of our model that we plan to address in the future is
that, if the true underlying process is less smooth in some regions
than others, or if there are spatial outliers, our model may
simultaneously both oversmooth \emph{and} undersmooth the random
effects, $\bZ$. In our gradient simulation in Section~\ref{SecSim},
the counties of Alameda (home of Oakland) and Solano have significantly
larger percentages of African Americans than any other county in the
state. As a result, the true underlying process that we've constructed
using (\ref{eqsimsetup}) for these counties takes much more extreme
values than their neighbors, resulting in oversmoothing in these
counties and creating the potential for undersmoothing in other
counties. While this issue is not unique to our model, this can lead to
poor estimation of the temporal gradients, such as biased estimates or
wide credible intervals. An approach similar to the spatially adaptive CAR (SACAR) model
proposed by \citet{ReiHod08} offers one possible solution:
replace the covariance matrix, $\Sigma_Z$, in~(\ref{EqSeparableCrossCov}) with
%
\begin{equation}
\Sigma_{Z}=R(\bphi)\otimes T(D-\alpha W)^{-1}T,
\label{eqsacar}
\end{equation}
where $T$ is a diagonal matrix with $T_{ii} = \sig_i$. We believe by
allowing each region to have its own variance parameter, outliers such
as Alameda and Solano in our simulation will receive larger $\sig_i$
(relative to the single variance parameter, $\sig$, described in this
paper) and, thus, will be less constrained by the magnitude of their
neighbors. Furthermore, regions which are more similar to their
neighbors would conceivably receive smaller $\sig_i$, allowing for
tighter credible intervals for both the random effects and their
gradients.\looseness=1

We certainly have not exhausted our modeling options from a theoretical
standpoint, either. Some of the richer association structures
described in Appendix~B of the online supplement may be appropriate in
alternate inferential contexts. While we demonstrated the advantages of
the process-based specifications over some simpler parametric options
for $Z_i(t)$ in our data analysis, one could envision alternative
specifications depending upon the inferential question at hand. For
example, if interest lay in separating the variability between time and
space using two variance parameters, additive specifications such as
$Z_i(t) = u_i + w(t)$, where $u_i$'s follow a Markov random field and
$w(t)$ is a temporal Gaussian process, could be explored. Now the
$u_i$'s and $w(t)$'s could have their own variance components. This,
however, would not allow the temporal functions to borrow strength
across the neighbors as effectively as we do here.

Apart from exploring such alternate specifications, our future work includes
expanding our focus to include spatiotemporal gradients for
point-referenced (geostatistical) data, where our response arises from
a spatiotemporal process $Y(\bs;t)$ with $\bs\in\Re^{d}$. Typically,
we have a finite collection of sites $\mathcal{ S}=\{\bs_1,\ldots
,\bs_n\}$ and time points $t\in\mathcal{ T} = \{t_{1},\ldots,t_{N_t}\}$
(as before) where the responses $Y(\bs_{i};t_{j})$ have been observed.
Spatiotemporal gradient analysis in this setting offers richer
possibilities, and of course avoids the problems associated with the
CAR model's failure to offer a true spatial process [Banerjee, Carlin and Gelfand
(\citeyear{BanCarGel04}), pages 82--83]. Here one can conceptualize spatial (directional)
gradients, temporal gradients or even ``mixed'' gradients.\looseness=1

\section*{Acknowledgments}
The authors are grateful to the NIH and the Air Resources Board of the
California
Environmental Protection Agency for providing the data and to the AE
and two referees whose comments greatly improved the paper.

\begin{supplement}[id=suppA]
\stitle{Imputation of missing daily hospitalization counts, MCMC
details, alternative models and comparison with discrete-time models\\}
\slink[doi]{10.1214/12-AOAS600SUPP} 
\sdatatype{.pdf}
\sfilename{aoas600\_supp.pdf}
\sdescription{As data for days with between one and four asthma
hospitalizations are missing, we impute county-specific values for
these days using a method similar to Besag's iterated conditional modes
method [\citet{Bes86}] but with means. We also lay out the details for
the MCMC implementation, discuss more general versions of our model and
compare our gradient estimates to finite differences from a simple
discrete-time model.}
\end{supplement}


%

\printaddresses

\end{document}